\pdfoutput=1

\documentclass[twocolumn]{autart}    

\usepackage{amsmath}
\usepackage{amsfonts}
\usepackage{amssymb}
\usepackage{graphicx}
\usepackage{tikz}
\usetikzlibrary{arrows,positioning}
\usetikzlibrary{calc}
\usetikzlibrary{decorations}
\usetikzlibrary{decorations.pathmorphing,%
decorations.fractals,%
decorations.shapes,%
decorations.text,%
decorations.pathreplacing,%
decorations.footprints,%
decorations.markings}

\usepackage{units}

\usepackage{nccmath}

\usepackage{subfig}

\usepackage[longnamesfirst,round]{natbib}

\begin{document}

\begin{frontmatter}
\title{Timing estimation in distributed sensor and control systems with central processing} 

\thanks{Corresponding author John-Olof Nilsson. Tel.: +46(0)739815787; E-mail address: jnil02@kth.se.}

\author{John-Olof Nilsson}\hspace{4mm}
\author{Peter H\"{a}ndel}

\address{Signal Processing Lab, ACCESS Linnaeus Centre, KTH Royal Institute of Technology, Osquldas v.10, 100\hspace{0.5mm}44 Stockholm, Sweden}

\begin{keyword}                           
Timing recovery; Clock synchronization; Distributed computer control systems; Time delay; Sensor fusion.               
\end{keyword}                             

\begin{abstract}                          
We consider the problem of estimating timing of measurements and actuation in distributed sensor and control systems with central processing. The focus is on direct timing estimation for scenarios where clock synchronization is not feasible or desirable. Models of the timing and central and peripheral time stamps are motivated and derived from underlying clock and communication delay definitions and models. Heuristics for constructing a system time is presented and it is outlined how the joint timing and the plant state estimation can be handled. For a simple set of underlying clock and communication delay models, inclusion of peripheral unit time stamps is shown to reduce jitter, and it is argued that in general it will give significant jitter reduction. Finally, a numerical example is given of a contemporary system design.
\end{abstract}

\end{frontmatter}

\section{Introduction}
Knowledge about timing is necessary in estimation and control of any dynamic system. Without the corresponding time instants, the state information, gained by a measurement, and the control capability, given by actuation, are limited. Consequently, measurement and control command timing must be measured (given time stamps) and estimated. This estimation task might seem trivial and indeed in tightly integrated systems, in which a central unit directly reads off measurements, performs processing, and controls actuation, estimating timing essentially entails setting equality between time stamps and estimates. Unfortunately, such centralized systems are often impractical in the sense that they limit the modularization, robustness, flexibility, and constrain the physical extent of the system. Instead in many applications, even though the processing is kept centralized, it is preferable to use distributed designs in which the sensing and actuation units are connected to a central unit by asynchronous communication links. This will improve robustness and flexibility and allow for higher modularization and indefinite physical extent of the system as each unit only need to be physically dependent on the remainder of the system through a communication link. Examples of such systems include popular networked control systems and sensor networks but also more traditional sensor and control systems with peripheral units connected by asynchronous data busses, e.g., USB.

Unfortunately, distributed systems give a number of timing-related difficulties. (See \citet{Lamport1978} for an excellent introduction to the perception of time in distributed systems.) The actuation and sensing itself is performed by units without a common time reference. Hence, time stamps from different units cannot be directly related to each other. The problem has been extensively studied in the computer and sensor network area, where many different clock synchronization strategies have been developed to create and maintain a common time reference. Surveys of the matter are found in \citet{Wu2011,Sivrikaya2004,Sundararaman2005}. However, rather flexible and capable peripheral units are required to implement clock synchronization. Also, the synchronization strategies consume communication resources, which might be limited or costly, and add tasks to the peripheral unit to perform for which it might have to interrupt its normal operation. As a result, clock synchronization will not be suitable for many sensor and control systems. 

If clock synchronization cannot be implemented or if it is undesirable for other reasons, timing must be estimated directly based on available measurements. In this article fundamental relations and models are derived from underlying clock and communication delay definitions and models, and heuristics are discussed for defining system time and handling coupling with the plant. Especially a naive approach, using only time stamps from a central unit, is contrasted with an approach using additional time stamps and information from the peripheral units. For simple underlying clock and communication delay models, the latter approach is shown to reduce jitter and it is argued that in general it can be expected to significantly reduce the jitter. Note that in this article \emph{jitter} refers to \emph{the difference between actual and estimated timing}.

Our experience is that many off-the-shelf sensor and actuation units do time stamp their actions, but clock synchronization cannot easily be implemented on them. That is, the motivation for studying the described estimation problem has come from experiences with sensor and control system implementations with off-the-shelf components. See also Section \ref{subsec:example} for an example.

The degrading effects on the performance of sensor and control systems caused by jitter are well known and minimizing jitter is of great importance in any such system \citep{Michiels2007}. To the first order, the effect of jitter is determined by the jitter multiplied by the dynamics of the plant or the control input \citep{Nilsson2010a,Shin1995}. Therefore, this work is believed to be especially relevant for networked control systems and sensor networks where communication delay variations can be substantial (large jitter) and for control and sensor systems with fast dynamics implemented with asynchronous data busses (small jitter but fast dynamics). However all distributed sensor and control systems will benefit from reduced jitter. The results presented in the article can also reduce system development from hardware or clock synchronization subsystems, hence facilitating development and system prototyping giving faster time-to-market.


It should be pointed out that clock synchronization and direct timing estimation are not in conflict and the techniques can be mixed for different peripheral units.

\subsection{Article overview}
The remainder of the article is organized as follows. In Section~\ref{sec:preliminaries}, the considered system setup is specified in detail and factors affecting the estimation problem structure are identified. In Section~\ref{subsec:previous_work}, a brief review of some previous works is given. In Section~\ref{sec:clock_delay_mod}, clock and communication delay definitions are given. In Section~\ref{sec:time_stamps}, measurement models for the time stamps are proposed and discussed. In Section~\ref{sec:time_reference}, a system time definition and implications thereof are discussed. In Section~\ref{sec:timing_dynamic}, additional relations are derived and further assumptions necessary for the estimation problem are made. In Section~\ref{sec:timing_model}, the final joint description of the timing and the time stamps is given. In Section~\ref{sec:heuristics}, difficulties and heuristics, for handling the coupling between timing and plant model, are discussed. In Section~\ref{sec:redu_jitter}, the jitter reduction, given by including peripheral unit time stamps and information, is quantified, and a numerical example is given. Finally, in Section~\ref{sec:conclusions}, conclusions of the article are drawn.

\section{Preliminaries}\label{sec:preliminaries}
The following distributed system setup is considered (refer to Figure~\ref{fig:scenario2} for an illustration):
The physical system consists of a \emph{central unit} $c$ interacting with a \emph{plant} via a set of \emph{peripheral units} $\{p_i\}$ in the form of \emph{actuators}, enabling actuation of the plant, and \emph{sensors}, providing measurements of the plant. The plant is assumed to be time invariant.
The peripheral units are connected to the central unit by asynchronous communication links. Each unit carries a separate clock $C_j:j\in\{c,i\}$ and the plant evolves according to physical time captured by the imaginary clock $C_r$. The relations between the clocks are unknown.
Timing instants $t^i_{k|r}$ (measurement and actuation instants) are given time stamps $\gamma^i_{k|c}$, indirectly by the central unit via the related communication, and potentially time stamps $\gamma^i_{k|i}$, directly by the peripheral unit.\footnote{Superscript $i$ indicates which peripheral unit the indexed quantity refers to; $k$ is a time index of timing instants of an individual peripheral unit; and subscript $|j:j\in\{c,i,r\}$ indicates which clock the indexed quantity is with respect to.} The desire is to estimate $t^i_{k|r}$. 
The number of actuators and sensors is arbitrary, but for the occurrence of the timing problem there has to be at least two peripheral units. 
Hence, the system description and the following analysis are applicable to both control and sensor systems.

\pgfdeclarelayer{background}
\pgfdeclarelayer{foreground}
\pgfsetlayers{background,main,foreground}
\tikzstyle{box}=[draw,fill=blue!15,minimum width=29mm,minimum height=11mm,text badly centered]
\tikzstyle{bbox}=[draw,fill=gray!25,minimum width=24mm,minimum height=15mm]

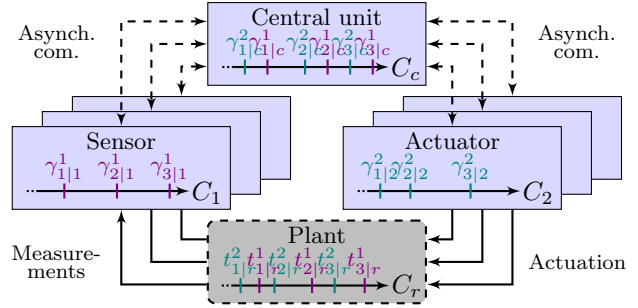
\begin{figure}[t!]
\centering
\begin{tikzpicture}[auto,>=latex']
    \node (cu) at (0,0) [box] {};
    \node (cu_text) at (cu) [above=1.25mm] {\small Central unit};

    \node (su) at (-26mm,-16.5mm) [box] {};
    \node (su_text) at (su) [above=1.25mm] {\small Sensor};

    \node (au) at (18mm,-16.5mm) [box] {};
    \node (au_text) at (au) [above=1.25mm] {\small Actuator};

    \node (plant) at (0,-29mm) [box,dashed,thick,rounded corners=1mm,fill=black!25] {};
    \node (plant_text) at (plant) [above=1.25mm] {\small Plant};

    \begin{pgfonlayer}{background}
        \node (su3) at (su.south west) [box,anchor=south west,above right=4mm and 8mm] {};
        \node (su2) at (su.south west) [box,anchor=south west,above right=2mm and 4mm] {};
        \node (au3) at (au.south west) [box,anchor=south west,above right=4mm and 8mm] {};
        \node (au2) at (au.south west) [box,anchor=south west,above right=2mm and 4mm] {};

        \draw[thick,->] (au.south) |- ($(plant.east)+(0,3mm)$);
        \draw[thick,->] (au2.south) |- ($(plant.east)+(0,0)$);
        \draw[thick,->] (au3.south) |- ($(plant.east)+(0,-3mm)$);

        \draw[thick,<-] (su.south) |- ($(plant.west)+(0,-3mm)$);
        \draw[thick,<-] (su2.south) |- ($(plant.west)+(0,0)$);
        \draw[thick,<-] (su3.south) |- ($(plant.west)+(0,3mm)$);
    \end{pgfonlayer}

    \draw[thick,<->,dashed] ($(cu.east)+(0,-3mm)$) -| (au.north);
    \draw[thick,<->,dashed] ($(cu.east)+(0,0mm)$) -| (au2.north);
    \draw[thick,<->,dashed] ($(cu.east)+(0,3mm)$) -| (au3.north);

    \draw[thick,<->,dashed] ($(cu.west)+(0,-3mm)$) -| (su3.north);
    \draw[thick,<->,dashed] ($(cu.west)+(0,0mm)$) -| (su2.north);
    \draw[thick,<->,dashed] ($(cu.west)+(0,3mm)$) -| (su.north);


    \node at ($(cu.west)+(-20mm,1.25mm)$) {\scriptsize Asynch.};
    \node at ($(cu.west)+(-20mm,-1.25mm)$) {\scriptsize com.};
    \node at ($(plant.west)+(-20mm,1.25mm)$) {\scriptsize Measure-};
    \node at ($(plant.west)+(-20mm,-1.25mm)$) {\scriptsize ments};
    \node at ($(cu.east)+(20mm,1.25mm)$) {\scriptsize Asynch.};
    \node at ($(cu.east)+(20mm,-1.25mm)$) {\scriptsize com.};
    \node at ($(plant.east)+(20mm,0)$) {\scriptsize Actuation};

    \draw[thick,->] ($(cu.south west)+(3.5mm,2.5mm)$) -> ($(cu.south east)+(-5mm,2.5mm)$) node[pos=1,anchor=west,below right=-2.5mm and -1mm] {$C_c$};
    \draw[thick,densely dotted,-] ($(cu.south west)+(2mm,2.5mm)$) -> +(1.5mm,0);
    \begin{scope}[violet]
    \draw[thick,-] ($(cu.south west)+(8mm,1.5mm)$) -> +(0,2mm) node[pos=1,anchor=south,above=-1.3mm] {\scriptsize $\gamma^1_{1|c}$};
    \draw[thick,-] ($(cu.south west)+(16mm,1.5mm)$) -> +(0,2mm) node[pos=1,anchor=south,above=-1.3mm] {\scriptsize $\gamma^1_{2|c}$};
    \draw[thick,-] ($(cu.south west)+(22mm,1.5mm)$) -> +(0,2mm) node[pos=1,anchor=south,above=-1.3mm] {\scriptsize $\gamma^1_{3|c}$};
    \end{scope}
    \begin{scope}[teal]
    \draw[thick,-] ($(cu.south west)+(5mm,1.5mm)$) -> +(0,2mm) node[pos=1,anchor=south,above=-1.3mm] {\scriptsize $\gamma^2_{1|c}$};
    \draw[thick,-] ($(cu.south west)+(13mm,1.5mm)$) -> +(0,2mm) node[pos=1,anchor=south,above=-1.3mm] {\scriptsize $\gamma^2_{2|c}$};
    \draw[thick,-] ($(cu.south west)+(19mm,1.5mm)$) -> +(0,2mm) node[pos=1,anchor=south,above=-1.3mm] {\scriptsize $\gamma^2_{3|c}$};
    \end{scope}

    \draw[thick,->] ($(su.south west)+(3.5mm,2.5mm)$) -> ($(su.south east)+(-5.5mm,2.5mm)$) node[pos=1,anchor=west,below right=-2.5mm and -1mm] {$C_{1}$};
    \draw[thick,densely dotted,-] ($(su.south west)+(2mm,2.5mm)$) -> +(1.5mm,0);
    \begin{scope}[violet]
    \draw[thick,-] ($(su.south west)+(7mm,1.5mm)$) -> +(0,2mm) node[pos=1,anchor=south,above=-1.3mm] {\scriptsize $\gamma^1_{1|1}$};
    \draw[thick,-] ($(su.south west)+(14mm,1.5mm)$) -> +(0,2mm) node[pos=1,anchor=south,above=-1.3mm] {\scriptsize $\gamma^1_{2|1}$};
    \draw[thick,-] ($(su.south west)+(21mm,1.5mm)$) -> +(0,2mm) node[pos=1,anchor=south,above=-1.3mm] {\scriptsize $\gamma^1_{3|1}$};
    \end{scope}

    \draw[thick,->] ($(au.south west)+(3.5mm,2.5mm)$) -> ($(au.south east)+(-5.5mm,2.5mm)$) node[pos=1,anchor=west,below right=-2.5mm and -1mm] {$C_{2}$};
    \draw[thick,densely dotted,-] ($(au.south west)+(2mm,2.5mm)$) -> +(1.5mm,0);
    \begin{scope}[teal]
    \draw[thick,-] ($(au.south west)+(5mm,1.5mm)$) -> +(0,2mm) node[pos=1,anchor=south,above=-1.3mm] {\scriptsize $\gamma^2_{1|2}$};
    \draw[thick,-] ($(au.south west)+(9mm,1.5mm)$) -> +(0,2mm) node[pos=1,anchor=south,above=-1.3mm] {\scriptsize $\gamma^2_{2|2}$};
    \draw[thick,-] ($(au.south west)+(17mm,1.5mm)$) -> +(0,2mm) node[pos=1,anchor=south,above=-1.3mm] {\scriptsize $\gamma^2_{3|2}$};
    \end{scope}

    \draw[thick,->] ($(plant.south west)+(3.5mm,2.5mm)$) -> ($(plant.south east)+(-5mm,2.5mm)$) node[pos=1,anchor=west,right=-1mm] {$C_r$};
    \draw[thick,densely dotted,-] ($(plant.south west)+(2mm,2.5mm)$) -> +(1.5mm,0);
    \begin{scope}[violet]
    \draw[thick,-] ($(plant.south west)+(7mm,1.5mm)$) -> +(0,2mm) node[pos=1,anchor=south,above=-1.3mm] {\scriptsize $\;\,t^1_{1|r}$};
    \draw[thick,-] ($(plant.south west)+(14mm,1.5mm)$) -> +(0,2mm) node[pos=1,anchor=south,above=-1.3mm] {\scriptsize $t^1_{2|r}$};
    \draw[thick,-] ($(plant.south west)+(21mm,1.5mm)$) -> +(0,2mm) node[pos=1,anchor=south,above=-1.3mm] {\scriptsize $t^1_{3|r}$};
    \end{scope}
    \begin{scope}[teal]
    \draw[thick,-] ($(plant.south west)+(4.5mm,1.5mm)$) -> +(0,2mm) node[pos=1,anchor=south,above=-1.3mm] {\scriptsize $t^2_{1|r}$};
    \draw[thick,-] ($(plant.south west)+(9mm,1.5mm)$) -> +(0,2mm) node[pos=1,anchor=south,above=-1.3mm] {\scriptsize $\quad t^2_{2|r}$};
    \draw[thick,-] ($(plant.south west)+(17mm,1.5mm)$) -> +(0,2mm) node[pos=1,anchor=south,above=-1.3mm] {\scriptsize $t^2_{3|r}$};
    \end{scope}

\end{tikzpicture}
\caption{The considered system set-up is composed of a central unit interacting with a plant via sensors and actuators communicating over asynchronous communication links. Each unit has its own perception of time from a local clock $C_j:j\in\{c,1,2\}$ and the actions in a peripheral units $i$ are given time stamps $\gamma^i_{k|i}$ and $\gamma^i_{k|c}$. The desire is to estimate the sampling instances $t^i_{k|r}$ based on measurements, actuation, and time stamps from all units.}\label{fig:scenario2}
\end{figure} 

The structure of the timing estimation problem is determined by a number of factors. Naturally, the main factor is the time stamps that are available or considered. For each peripheral unit, there might be time stamps from the central unit only or time stamps from both the central unit and the peripheral unit. These two situations will be referred to as \emph{scenario 1} and \emph{scenario 2}. 

\emph{Scenario 1}: Only time stamps from the central unit are available for each timing instant. This means that all time stamps are with respect to the same clock, but unknown communication delays give unknown delays in the relation between time stamps and the corresponding time instants. Hence the relation between time stamps and timing instants are governed by models for the communication delays.

\emph{Scenario 2}: In addition to the central unit time stamps, time stamps from the peripheral units are available. The relation between these time stamps and the corresponding sampling instants are not affected by communication delays and are therefore in some sense of higher quality. However, instead of the unknown delays there are unknown clock relations. These unknown clock relations are governed by models for the clocks which now appear in addition to the communication delay models. It is clear that in comparison with scenario 1 there is more information available but how to exploit it is not obvious.

In addition to the available time stamps, the triggering modes of the actions can also add structure and information. The actions in the peripheral units can be \emph{command-triggered}, \emph{time-triggered}, \emph{event-triggered}, or \emph{triggered at random}. The triggering modes add time information by constraining the timing by causality (action must take place after a command is sent and before a report/measurement is received), by adding relative time information between timing instants (regular intervals of time-triggering), and by restraining the plant-state-to-timing-instant relation (constraint of triggering events). 
Time-triggering provides similar information to that of peripheral unit time stamps. Therefore the scenario with only central unit time stamps and time-triggering will be considered belonging to scenario 2. See Section~\ref{sec:timing_dynamic} for further discussion about this. 

\section{Previous work}\label{subsec:previous_work}
A large number of publications dealing with unknown possibly time-varying (communication) delays in sensor and control systems can be found. For specific communication delay models, these problems are equivalent with scenario 1. Surveys of the matter with a focus on control and transfer functions are found in~\citet{Ferreira1997,Bjoerklund2003}. Treatments of the problem with a stronger focus on estimation and state space descriptions can be found in~\citet{Julier2005,Nilsson2010a,Thomopoulos1994a}. 

In contrast, publications regarding scenario 2 are relatively scarce. In~\citet{Philipp2010}, the problem of estimating the timing of sensor measurements in a control system is based on time stamps from both the central and the sensor units. The estimation is treated jointly with the system state and the system parameter estimation by constraining the relation between the central and the sensor unit clocks to be affine over a time window. Based on heuristic arguments, a model, corresponding to the special case of time-triggered sensors, has earlier been presented in~\citet{Nilsson2010}. 

Regarding general modeling of time stamps, clocks, and time in distributed systems many related studies can be found in the clock synchronization literature \citep[e.g.,][]{Wu2011,Sivrikaya2004,Sundararaman2005} and in the instrumentation and measurement literature \citep[e.g.,][]{Galleani2003,Audoin1993,Paxson1997}.

\section{Clocks and communication delays}\label{sec:clock_delay_mod}
The time stamps in the system are taken from the local clocks in the different units. In turn, the time between a time instant in one unit and a related time stamp in another unit are determined by communication delays. Consequently, the relation between the time stamps and the timing will be governed by clocks and communication delays.

\subsection{Clock definitions}\label{sec:system_clocks}
The perception of time in the system, and therefore also the timing, is given by clocks. A clock $C_j$ is a device used to measure time $t$. The reading of clock $C_j$ at time $t$ is denoted with
\begin{equation}\label{eq:clock_reading}
t_{|j}=C_j(t).
\end{equation}
$C_j(t)$ is assumed to be a continuous and differentiable function. A discrete clock progressing in ``ticks'' can be thought of as a continuous clock with up to $\nicefrac{1}{2}$ ``tick'' error in the reading~\citep{Lamport1978}.

The time, as perceived by the physical plant, is given by the imaginary clock $C_r$. For simplicity, it will be assumed that
\begin{equation}\label{eq:phys_time}
t_{|r}=C_r(t)=t
\end{equation}
and therefore the subscript $|r$ will be dropped for time instants and time differentials given with respect to $C_r$. However, to emphasize the clock interpretation of time as perceived by the plant, $C_r(t)$ will often be used instead of $t$. \emph{Mutatis mutandis}, the analysis could be changed to handle arbitrary clock relations instead of \eqref{eq:phys_time}.

The \emph{clock off-set} $\beta^j_{|r}(t)$ of a clock $C_j$, with respect to the clock $C_r$, is the momentary difference between the clock readings,
\begin{equation}\label{eq:clock_offset}
\beta^j_{|r}(t)=C_j(t)-u^j_{|r}C_r(t)
\end{equation}
where $u^j_{|r}$ is a dimensionless time unit conversion factor. The \emph{clock skew} $\stackrel{\angle}{\alpha}\!{}^j_{|r}(t)$ is the time derivative of the clock offset,
\begin{equation}\label{eq:clock_skew}
\stackrel{\angle}{\alpha}\!{}^j_{|r}(t)=\frac{d\beta^j_{|r}(t)}{dt}.
\end{equation}
The second-order derivative will be referred to as \emph{clock drift}.
These clock definitions are consistent with those found in the clock synchronization literature~\citep{Wu2011,Sundararaman2005}. 

From~\eqref{eq:phys_time}, \eqref{eq:clock_offset}, and~\eqref{eq:clock_skew}, the clock reading $C_j(t)$ is given by
\begin{equation}\label{eq:clock_model_cont}
t_{|j}=C_j(t)=\int_{t_0}^t\alpha^j_{|r}(\tau)d\tau + \beta^j_{|r}(t_0) +u^j_{|r}t_0
\end{equation}
where $t_0$ is some arbitrary time origin and $\alpha^j_{|r}(t)$ is the \emph{clock pace} defined by $\alpha^j_{|r}(t)=\stackrel{\angle}{\alpha}\!{}^j_{|r}(t)+u^j_{|r}$.

\begin{figure}[tb!]
\centering
\begin{tikzpicture}[auto,>=latex']
\draw[->, very thick] (-5mm,0) -- (65mm,0) node[below right] {$t$};
\draw[->, very thick] (0,-5mm) -- (0,50mm) node[left=1mm] {$\nicefrac{t_{|j}}{t_{|r}}$};

\draw[-,very thin,dashed] (0,0) -- (65mm,50mm) node[pos=0.835,dotted,pin={[dotted,pin distance=3mm,pin edge={<-,shorten <=1mm,shorten >=-0.5mm}]-65:$C_r(t)$}] {};
\draw[-,thick,decorate,decoration={random steps,segment length=3mm}] (0,5mm) -- (22mm,15mm) -- (33mm,21mm) -- (60mm,50mm) node[pos=0.75,pin={[pin distance=0mm,pin edge={<-,shorten <=-1.5mm,shorten >=-1mm}]135:$C_{j}(t)$}] {};

\draw[|-|] (-7mm,0) -- (-7mm,5mm) node[midway,left,anchor=east] {$\beta^{j}_{0|r}$};
\draw[-,dotted] (-7mm,5mm) -- (0,5mm);

\node (alpha1) at (22mm,15mm) [circle,fill,minimum size=1mm,inner sep=0] {};
\node (alpha2) at (33mm,21mm) [circle,fill,minimum size=1mm,inner sep=0] {};
\draw[|-|] (22mm,11mm) -- (33mm,11mm) node[pos=0.5,anchor=north] {$dt^{j}_{k}$};
\draw[|-|] (37mm,15mm) -- (37mm,21mm) node[pos=0.5,anchor=west] {$\alpha^{j}_{k|r} dt^{j}_{k}$};
\draw[-,dotted] (22mm,0mm) -- (22mm,15mm);
\draw[-,dotted] (33mm,0mm) -- (33mm,21mm);
\draw[-,dotted] (37mm,15mm) -- (0mm,15mm);
\draw[-,dotted] (37mm,21mm) -- (0mm,21mm);
\node (tkm1) at (23mm,0mm) [label={[label distance=-1mm]below:$t^j_{k-1}$}] {};
\node (tk) at (33mm,0mm) [label={[label distance=-1mm]below:$t^j_{k}$}] {};
\node (tkm1j) at (0mm,15mm) [label={[label distance=-1mm]left:$t^j_{k-1|j}$}] {};
\node (tkj) at (0mm,21mm) [label={[label distance=-1mm]left:$t^j_{k|j}$}] {};

\node (t0) at (0,0) [pin={[pin distance=1.5mm,pin edge={<-,shorten >=-1mm}]below right:$t_0$}] {};
\end{tikzpicture}
\caption{Illustration of clock readings and clock model parameters of a clock $C_j$ with respect to the imaginary clock $C_r$. 
}\label{fig:clock}
\end{figure}
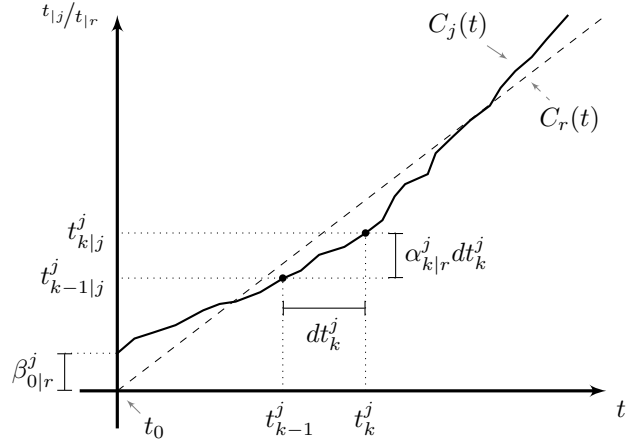

For a discrete time system, clock readings are only relevant at discrete time instants. 
At the time instants $t^{\gamma_{i,j}}_{k}$ the corresponding clock reading, i.e., time stamps $\gamma^j_{k|j}$, are recorded,
\begin{equation}\label{eq:ts_and_ti_conv}
\gamma^i_{k|j}=C_j(t^{\gamma_{i,j}}_{k}),
\end{equation}
where as the timing instants $t^{i}_{k}$ are to be estimated.
For an arbitrary set of ordered instants $\{t^i_{k}\}$, \eqref{eq:clock_model_cont} can be expressed as
\begin{equation}
t^i_{k|j}=C_j(t^i_{k})=\medop\sum_{l=1}^{k}\alpha^j_{l|r} dt^i_{l} + \beta^j_{0|r}\label{eq:clock_model_disc}
\end{equation}
where $dt^i_{l}=t^i_{l}-t^i_{l-1}$, $\beta^j_{0|r}=\beta^j_{|r}(t^i_{0})+u^j_{|r}t^i_0$, and $\alpha^j_{l|r}$ is the mean clock pace over the interval $[t^i_{l},t^i_{l-1}]$,
\begin{equation*}
\alpha^j_{l|r}=\frac{1}{dt^i_{l}}\int_{t^i_{l-1}}^{t^i_{l}}\alpha^j_{|r}(\tau)d\tau.
\end{equation*}
As only discrete time instants are considered the mean clock pace over intervals will simply be referred to as the \emph{clock pace}. The clock reading and the clock parameters are illustrated in Figure~\ref{fig:clock}.

\subsection{Communication delay definitions}\label{sec:com_char}
\begin{figure}[tb!]
\centering
\begin{tikzpicture}[auto,>=latex']
\draw[->, very thick] (-5mm,0) -- (70mm,0) node[below right] {$k$};
\draw[->, very thick] (0,-5mm) -- (0,50mm) node[left=1mm] {$t$};

\draw[-,thick] (0,25mm) -- (2.5mm,17mm) -- (5mm,23mm) -- (7.5mm,35mm) -- (10mm,28mm) -- (12.5mm,51mm) -- (15mm,35mm) -- (17.5mm,41mm) -- (20mm,33mm) -- (22.5mm,45mm) -- (25mm,33mm) -- (27.5mm,25mm) -- (30mm,31mm) -- (35mm,14mm) -- (37.5mm,25mm) -- (40mm,22mm) -- (42.5mm,33mm) -- (45mm,31mm) -- (47.5mm,23mm) -- (50mm,17mm) -- (52.5mm,21mm) -- (55mm,15mm) -- (57.5mm,20mm) -- (60mm,18mm) -- (62.5mm,23mm) -- (65mm,22mm);
\draw[-,dashed, thin] (0,27mm) -- (2.5mm,28mm) -- (5mm,29mm) -- (7.5mm,31mm) -- (10mm,32mm) -- (12.5mm,33mm) -- (15mm,33.5mm) -- (17.5mm,33.5mm) -- (20mm,32.5mm) -- (22.5mm,31mm) -- (25mm,29mm) -- (27.5mm,27mm) -- (30mm,25mm) -- (35mm,23mm) -- (37.5mm,23mm) -- (40mm,22.5mm) -- (42.5mm,22.5mm) -- (45mm,22mm) -- (47.5mm,22mm) -- (50mm,21mm) -- (52.5mm,20mm) -- (55mm,19mm) -- (57.5mm,18mm) -- (60mm,18mm) -- (62.5mm,18mm) -- (65mm,18mm);

\node at (65mm,18mm) [pin={[pin distance=1.5mm,pin edge={<-,shorten >=-1mm}]0:$\eta^i_k$}] {};
\node at (65mm,22mm) [pin={[pin distance=1.5mm,pin edge={<-,shorten >=-1mm}]20:$\delta t^i_k$}] {};
\end{tikzpicture}
\caption{Illustration of the dynamics of a communication delay $\delta t^i_k$ and the underlying trend captured in the state $\eta^i_k$.}\label{fig:com_delay}
\end{figure}
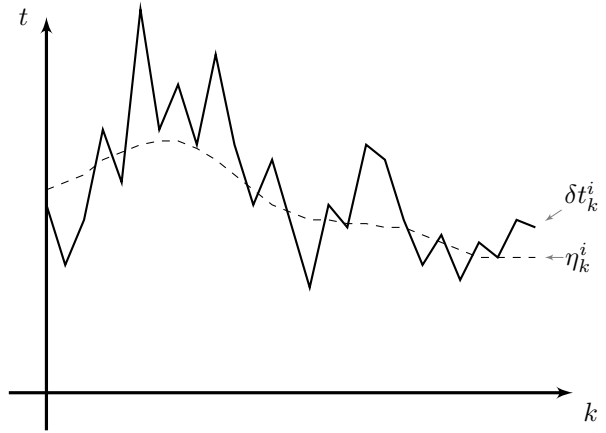

The asynchronous communication links transfer the commands and the reports between the central and the peripheral units. Because these messages are transmitted in response to or with a request of an action, the communication itself carries information about the timing. Assuming error-free communication, from an external point of view, the communication link only has the effect of delaying messages. 
Hence, for the purposes of timing estimation, the communication link is described by a communication delay. For simplicity, the communication delay $\delta t^i_k$ will be defined such that it includes all delays at time instant $t^i_k$ in communicating that a measurement has been taken, or in getting a control command executed, in the peripheral unit $p_i$, meaning that
\begin{equation}\label{eq:com_delay}
t^{\gamma_{i,c}}_{k}=\delta t^i_k+t^i_{k}.
\end{equation}
The communication delay in turn is divided into a communication delay state term $\eta^i_k$ and a white stochastic residual term $\nu^i_k$,
\begin{equation}\label{eq:div_innov}
\delta t^i_{k}=\eta^i_k+\nu^i_k.
\end{equation}
The dynamics of the communication delay, and the underlying trend captured in the state term, are illustrated in Figure~\ref{fig:com_delay}.

\section{Time stamps}\label{sec:time_stamps}
The time stamps are the primary measurements based on which timing is to be estimated. For this purpose, measurement models are needed. The time stamps from the central unit are taken at transmission of control commands to, and reception of reports from, the peripheral units. In the command-triggered case there might be two time stamps, while in the remaining triggering modes there will be a single time stamp. In this analysis, the related quantities are distinguished by a second superscript $t$ (transmit) and $r$ (receive), respectively.
The measurement models of the time stamps from the central unit are given by the following proposition:
\begin{prop}\label{lem:c_ts}
The relation between the time stamps from the central unit and the related timing instant $t^{i}_{k}$ can be taken to have the form
\begin{equation}\label{eq:lemma1}
\gamma^{i}_{k|c}=C_c(t^{i}_{k}+m^i_k+v^i_k)
\end{equation}
where $\gamma^{i}_{k|c}$ is (the potentially weighted arithmetic mean of) the considered time stamp(s) and, depending on the action control mode, $m^i_k$ is the positive or the negative delay state or a bounded difference of the reception or transmission delay states and where $v^i_k$ is a white stochastic residual.
\end{prop}
If a report is received from a peripheral unit then from \eqref{eq:com_delay} and \eqref{eq:div_innov}
\begin{equation}\label{eq:ts_received_meas}
t^{\gamma_{i,j},r}_{k}=t^{i}_{k}+\eta^{i,r}_k+\nu^{i,r}_k
\end{equation}
where by causality $\eta^{i,r}_k>0$ and $\eta^{i,r}_k+\nu^{i,r}_k>0$. \emph{Vice versa}, if a command is transmitted
\begin{equation}\label{eq:ts_transmit_com}
t^{i}_{k}=t^{\gamma_{i,c},t}_{k}+\eta^{i,t}_k+\nu^{i,t}_k
\end{equation}
where $\eta^{i,t}_k>0$ and $\eta^{i,t}_k+\nu^{i,t}_k>0$.
From \eqref{eq:ts_and_ti_conv}, relations \eqref{eq:ts_received_meas} and \eqref{eq:ts_transmit_com} can be written in the form \eqref{eq:lemma1}
where $\gamma^{i}_{k|c}$ is the available time stamp, $m^i_k$ and $v^i_k$ have the obvious meanings and $m^i_k$ and $m^i_k+v^i_k$ are constrained to be positive or negative, respectively. 

If a command is transmitted and subsequently a report is received, there are both a command transmission time stamp $\gamma^{i,t}_{k|c}$ and a report reception time stamp $\gamma^{i,r}_{k|c}$ available from the central unit. These two time stamps could potentially be treated independently giving two instances of \eqref{eq:lemma1}. As an alternative either one could be ignored, giving \eqref{eq:lemma1} again, or the (potentially weighted) arithmetic mean of the time stamps can be used. From \eqref{eq:ts_received_meas} and \eqref{eq:ts_transmit_com}
\begin{align}
\bar{t}^{\gamma_{i,c}}_{k}=&(t^{\gamma_{i,c},t}_{k}+t^{\gamma_{i,c},r}_{k})/2\nonumber\\
            =&t^{i}_{k}+\frac{\eta^{i,t}_k-\eta^{i,r}_k}{2}+\frac{\nu^{i,t}_k-\nu^{i,r}_k}{2}\nonumber\\
            =&t^{i}_{k}+\eta^{i,tr}_k+\nu^{i,tr}_k\label{eq:ts_tr_mean}
\end{align}
where $\eta^{i,tr}_k$ and $\nu^{i,tr}_k$ has the obvious meanings. 
The relation \eqref{eq:ts_tr_mean} can be written on the form \eqref{eq:lemma1}
where $\gamma^{c}_{k|c}=(\gamma^{c,t}_{k|c}+\gamma^{c,r}_{k|c})/2$, $|m^{i,tr}_k|<\alpha^{i,-1}_{k|r}(\gamma^{c,r}_{k|r}-\gamma^{c,t}_{k|r})/2$, $|m^{i,tr}_k+v^{i,tr}_k|<{\alpha^{i,-1}_{k|r}}^{-1}(\gamma^{i,r}_{k|r}-\gamma^{i,t}_{k|r})/2$ and the terms $m^i_k$ and $v^i_k$ contain small contributions from the clock drift. Due to clock stability these contributions can normally be ignored. (See Section \ref{sec:time_reference} for further discussion about this.)
Note that unless $\nu^{i,t}_k$ and $\nu^{i,r}_k$ are Gaussian, there will be some information loss in creating the arithmetic mean. However, it is unlikely that such detailed knowledge about the statistics of the residual terms is available such that the loss is significant.


The measurement models of the peripheral unit time stamps are given by the following proposition:
\begin{prop}\label{lem:p_ts}
The  relation between the time stamps $\gamma^{i}_{k|i}$ and the related timing instants $t^{i}_{k|r}$ is
\begin{equation}\label{eq:peripheral_time_stamps}
\gamma^{i}_{k|i}=C_{i}(t^{i}_{k})+q^{i}_{k}
\end{equation}
where $q^i_{k}$ is an unknown delay between the time stamping and the timing instant with respect to the local clock $C_{i}$.
\end{prop}
The relation is a direct consequence of the clock definition and the time-stamping procedure. With respect to the local clock $C_i$,
\begin{equation*}
\gamma^{i}_{k|i}=t^{i}_{k|i}+q^{i}_{k}.
\end{equation*}
From \eqref{eq:ts_and_ti_conv}, the relation \eqref{eq:peripheral_time_stamps} is attained.

The time stamps from the central unit and the time stamps from the peripheral units are fundamentally different in the sense that the former is affected by communication delays potentially affected by external factors, while the latter is only affected by internal factors. This is the reason the delay $m^i_k+v^i_k$ is expressed with respect to the ``external" clock $C_r$ while the delay $q^i_{k}$ is expressed with respect to $C_{i}$.

\section{System time}\label{sec:time_reference}
In general, time will be defined by the observed plant. Prior knowledge about and possible explicit time variations of the plant will be given with respect to some time, providing an absolute time reference. In case the plant model is time invariant, a dynamic model (essentially unit and constant definitions) still determines the pace of time (e.g., a mass and Newton's second law of motion). Together, this would give the physical time $C_r$ of the system and, given available time stamps and plant measurements and control commands collected in the sets $\Gamma_k$ and $D_k$ respectively, the timing instants could in principle be estimated by the conditional expectations $E(t^i_{k}|\Gamma_k,D_k)$. However, this is an idealized view and in most cases an unambiguous physical time of the plant cannot be based on prior knowledge and model assumptions. 
Consequently, a suitable system time needs to be constructed.

From \eqref{eq:clock_model_disc} the time, as perceived by a clock, is defined by an off-set and a clock pace. 
The system has been assumed time invariant, and the time of the initial state of the system is often either static or specified in relation to the first measurement making off-set relative $C_r$ arbitrary. Thus, a reference clock in the system must be chosen, against which the off-set can be set. For control systems the controller need to be able to predict the effect of a control command and therefore the reference clock is suitably taken to be the central unit clock $C_c$. For notational simplicity, the readings of the reference clock are assumed to have the same units as that of $C_r$, i.e., $u^c_{|r}=1$. It might be tempting to set $C_c(t)=C_r(t)$ defining both off-set and clock pace. However, as will be shown later, this is a poor choice as it will lead to an unobservable system and give little insight into the limits of the perception of time.

The time invariance of the plant also means that a constant communication delay can only be observed relative to the delay of another peripheral unit via plant measurements or plant actuation. Hence, a reference peripheral unit $s$ needs to be chosen, which means that the timing estimates of that reference peripheral unit are assumed independent of all plant measurements and control commands.
This implies that all other communication delays will be estimated relative to the timing estimates of the reference peripheral unit. The off-set between $C_c$ and $C_r$ is then
suitably defined by setting the reference peripheral unit timing estimates in the reference clock time frame, from the perspective of the remainder of the system, to its conditional expected values
\begin{equation}
t^s_{k|c}\Rightarrow E(t^s_{k|c}|\Gamma_k).\label{eq:time_ref}
\end{equation}
As the time stamps, related to different peripheral units, are only coupled via the plant measurements and actuation, this means that $t^s_{k|\varsigma}$ is separately estimated and subsequently, from the perspective of the remainder of the system, considered perfectly known.
From the definition \eqref{eq:clock_reading} and \eqref{eq:clock_offset}
\begin{align*}
t^s_{k|c}&=C_c(t^s_k)\\
                 &=C_r(t^s_k)+\beta^c_{r}(t^s_k)\\
                 &=t^s_{k}+\beta^c_{r}(t^s_k)
\end{align*}
and therefore \eqref{eq:time_ref} is equivalent with
\begin{equation}\label{eq:ref_clock}
C_c(t^s_k)=C_r(t^s_k)+E(t^s_{k|c}|\Gamma_k)-t^s_{k}.
\end{equation}
As for the clock pace, 
a nonzero clock skew gives scale errors in the time differentials. The frequency stability of clock oscillators is typically well below 1\nicefrac{0}{00} (typically in the range 100-10ppm). This means that the induced errors in a plant state over a time period would be well below 1$\nicefrac{0}{00}$ of the change in the state. The observability of such errors would normally be, at best, weak. In case they could be estimated, the interpretation of this would be that the plant itself would work as a clock with comparable quality as the reference clock, which in most cases would be absurd. However, any time variation of the terms $E(t^s_{k|c}|\Gamma_k)-t^s_k$ in \eqref{eq:ref_clock} will be perceived as a nonzero skew in the reference clock. This gives equivalent scale errors, which could potentially be estimated by its coupling with the plant model. On the other hand, assuming the communication delay bounded, the longer the correlation of the delay, giving more measurements to estimate the scale errors, the smaller the scale errors must be, making them more difficult to estimate. In other words, the signal-to-noise ratio is inherently poor. Therefore, in general, it will be assumed that the reference clock pace is known and for notational simplicity it is assumed that $\alpha^c_{|r}(t)=1$. This implies that
\begin{equation}\label{eq:const_offset}
\beta^c_{0|r}=E(t^s_{k|c}|\Gamma_k)-t^s_k
\end{equation}
where $\beta^s_{0|r}$ is a constant, and that \eqref{eq:ref_clock} is valid for all time instants,
\begin{equation*}
C_c(t)=C_r(t)+\beta^c_{0|r}.
\end{equation*}

Note that the terms $E(t^s_{k|c}|\Gamma_k)\,-\, t^s_k$ in \eqref{eq:ref_clock} are also equivalent with $E(\delta t^s_{k|c}|\Gamma_k)-\delta t^s_k$. This means that the stability of our time reference will be limited by how well the time variation of the delay can be estimated. Naturally, without any well characterized connection to the plant, our perception of the timing of actions taken at the plant will be poor. This suggests that preferably the reference peripheral unit should have a stable connection to the central unit.

The assumptions about the system time are summarized in the following:
\begin{assum}\label{def:time}
System time is defined in relation to $C_c$ and a reference peripheral unit $s$ by
\begin{equation*}
t^s_{k|c}\Rightarrow E(t^s_{k|c}|\Gamma_k)\quad\text{and}\quad\alpha^c_{|r}(t)=1
\end{equation*}
giving
\begin{equation*}
C_c(t)=C_r(t)+\beta^c_{0|r}
\end{equation*}
where $\beta^c_{0|r}=E(t^s_{k|c}|\Gamma_k)-t^s_{k}$. 
\end{assum}

\begin{cor}\label{lem:rel_delay}
Proposition \ref{lem:c_ts} and Assumption \ref{def:time} imply
\begin{equation}\label{eq:lem_rel_delay}
\gamma^i_{k|c}=t^{i}_{k|r}+\delta^i_k+v^i_k
\end{equation}
where the mean relative delay $\delta^i_k=m^i_k-\beta^c_{0|r}$.
\end{cor}


\begin{cor}\label{cor:ref_unit_meas}
Corollary~\ref{lem:rel_delay} and Assumption~\ref{def:time} imply
\begin{equation}\label{eq:closed_form}
\gamma^s_{k|c}=t^s_{k|r}+v^s_k.
\end{equation}
\end{cor}
Corollary \ref{cor:ref_unit_meas} makes it clear why setting $C_\varsigma(t)=C_r(t)$ would be a poor choice. With this choice \eqref{eq:lem_rel_delay} would still be of the same form, but \eqref{eq:closed_form} would not hold true. There would be an unknown delay for the reference unit but as previously argued these delays are only observable relative to each other so all delays would not be simultaneously observable in general.

The equation \eqref{eq:lem_rel_delay} is deceptive in the sense that $\beta^c_{0|r}$ is unknown. Even if the timing instants $t^{i}_{k}$ and the mean relative delays $\delta^i_k$ were observable, only the relative timing is observable. Fortunately, as $\beta^c_{0|r}$ is arbitrary with respect to the plant, this does not matter, but if a measurement of the system with zero delay became available, this would not mean that the time relation to other measurements and actuation would be known. For this, we would need to know $\beta^c_{0|r}$. The effect of \eqref{eq:const_offset} is that an unknown, in the communication with the peripheral unit $s$, is hidden in the reference clock offset.



In case the assumption $\alpha^s_{|r}(t)=constant$ cannot be made or one wishes to model associated errors explicitly, $\alpha^s_{k|r}$ can be added as a plant state and inserted as an unknown correcting factor for time differentials.


\section{Timing dynamics}\label{sec:timing_dynamic}
These far models of individual time stamps have been proposed, but for a single timing instant the number of unknowns is larger than that of time stamps. Therefore, some additional relations are needed expressing dependencies between different timing instants. The combination of these relations is referred to as the \emph{timing dynamics}.

Peripheral unit time stamps or time-triggering give information about the temporal relation between timing instants related to the same peripheral unit. The corresponding timing dynamics relations are described in the following proposition:
\begin{prop}\label{lem:timing_dynamic}
For scenario 2 there is the following timing dynamics relation
\begin{equation}\label{eq:timing_dynamic}
t^{i}_{k+1}=\alpha^{i,-1}_{k+1|r}d^{i}_k+t^{i}_{k}+w^i_{k|q}
\end{equation}
where $d^{i}_k=\gamma^{i}_{k+1|c}-\gamma^{i}_{k|c}$ or $d^i_k$ is an unknown but repeating time period of the time-triggering and the noise term $w^i_{k|q}$ is the uncertainty about the time period.
\end{prop}

From \eqref{eq:clock_model_disc}
\begin{equation}\label{eq:time_diff_skew}
C_i(t^{i}_{k+1})-C_i(t^{i}_{k})=\alpha^i_{k+1|r}(t^{i}_{k+1}-t^{i}_{k})
\end{equation}
and from Proposition \ref{lem:p_ts}
\begin{equation}\label{eq:diff_perf_ts}
C_i(t^{i}_{k+1})-C_i(t^{i}_{k})=d^{i}_{k}+dq^i_{k}
\end{equation}
where $dq^i_{k}=q^i_{k+1}-q^i_{k}$. 
The same relation also holds in case there is no peripheral unit time stamps, but the process is time-triggered (regular interval). In this case, $d^{i}_{k}$ will be interpreted as one out of possible multiple unknown intervals and $dq^i_{k}$ as some possible randomness associated with the intervals.

Together, \eqref{eq:time_diff_skew} and \eqref{eq:diff_perf_ts} give the timing dynamics relation \eqref{eq:timing_dynamic}
where the noise term $w^i_{k|q}=\alpha^{i,-1}_{k|c}dq^i_{k}$.

Note that $q^{i}_{k}$ is a constant in case the time-stamping process is deterministic with respect to the number of clock cycles in relation to the action in the peripheral unit. This means that $dq^i_k=0$, implying $w^i_{k|q}=0$, and such a delay does not matter. $dq^i_k$ will be assumed unknown but with known statistics. In general a dynamic model for $dq^i_k$ could be used.

From Proposition \ref{lem:timing_dynamic} and Corollary \ref{lem:rel_delay} it is seen that the dynamic descriptions of $\delta^i_k$ and $\alpha^{i,-1}_{k|r}$ are a part of the timing dynamics. Accordingly, the following assumptions are made:

\begin{assum}\label{assum:com_del_descr}
For scenarios 1 and 2, a dynamic model for the mean relative delay $\delta^i_k$ is given, possibly parameterized with additional parameters $\vartheta^i_k$ and with a driving noise $w^i_{k|\delta}$,
\begin{equation}\label{eq:dyn_mod_comdelay}
[\delta^i_{k+1},\vartheta^i_{k+1}]=g_i(\delta^i_{k},\vartheta^i_{k},w^i_{k|\delta}).
\end{equation}
\end{assum}
Apart from $\beta^\varsigma_{0|r}$, the dynamic model \eqref{eq:dyn_mod_comdelay} is essentially a model of the communication delay. However, in the case of command-triggering it can be a model for the asymmetry of the communication delay. Examples of dynamic models of communication delays is rather scarce in the literature. Some examples of models and detailed analysis of underlying mechanisms can be found in~\citet{Myakotnykh2010,Paxson1997,Beuerman1988,Ganeriwal2003,Maroti2004}.

\begin{assum}\label{assum:inv_clock}
For scenario 2, a dynamic model for the inverse clock pace $\alpha^{j,-1}_{k|r}$ is given, possibly parameterized with additional states $\theta^j_{k|r}$ and with a driving noise $w^j_{k|\alpha}$,
\begin{equation*}
[\alpha^{j,-1}_{k+1|r},\theta^j_{k+1|r}]^T=f_j(\alpha^{j,-1}_{k|r},\theta^j_{k|r},w^j_{k|\alpha}).
\end{equation*}
\end{assum}
Unfortunately typically models for the clock pace, and not the inverse of the clock pace, are found in the literature. Fortunately, the stability of most clocks will let us approximate many models in terms of the inverse clock.
Examples of clock models and characteristics are found in~\citet{Galleani2003,Zucca2005,Audoin1993}.

\section{Timing and time stamps}\label{sec:timing_model}
The timing and its relation to the time stamps are described by the following proposition:
\begin{prop}\label{prop:final_model}
For scenarios 1 and 2, Corollary \ref{lem:rel_delay} and Assumption \ref{assum:com_del_descr} give
\begin{align}
[\delta^i_{k+1},\vartheta^i_{k+1}]&=g_i(\delta^i_{k},\vartheta^i_{k},w^i_{k|\delta})\label{eq:3}\\
\gamma^i_{k|c}&=t^{i}_{k|r}+\delta^i_k+v^i_k.\label{eq:4}
\end{align}
For scenario 2, Proposition \ref{lem:timing_dynamic} and Assumption \ref{assum:inv_clock} additionally give
\begin{align}
t^{i}_{k+1}&=\alpha^{i,-1}_{k|r}d^{i}_k+t^{i}_{k}+w^i_{k|q}\label{eq:1}\\
[\alpha^{i,-1}_{k+1|r},\theta^i_{k+1|c}]&=f_i({\alpha^i_{k|r}}^{-1},\theta^i_{n|r},w^i_{k|\alpha})\label{eq:2}
\end{align}
From Corollary \ref{cor:ref_unit_meas}, for the reference unit $s$, equation \eqref{eq:4} becomes $\gamma^s_{k|c}=t^s_{k|r}+v^s_k$ and model \eqref{eq:3} is superfluous.
The timing of the complete systems is described by one instance of \eqref{eq:3}-\eqref{eq:2} for each peripheral unit $p_i$.
\end{prop}


In case the peripheral unit is time-triggered extra states might be added for the possibly unknown sampling periods. Note that the process noise $w^i_{k|q}=\alpha^{i,-1}_{k|c}dq^i_{k}$ is dependent on the clock pace. However, due to clock stability this coupling does not normally have to be modeled explicitly. Also the noise terms $w^i_{k|\alpha}$ and $w^i_{k|q}$ would typically scale with $d^{i}_k$ but similarly this does not normally have to be modeled explicitly. 

\section{Plant coupling heuristics}\label{sec:heuristics}

Based on time stamps, plant measurements, and plant actuation, we wish to estimate the plant state. In doing so, the timing must be estimated. Obviously, the systems \eqref{eq:3}-\eqref{eq:2} are not in general independently observable without considering the coupling with a plant model. This means that the coupling between the time instants and the plant must be modeled unless some additional information is available, such as external communication delay measurements, or some additional assumptions are made, such that the communication link is symmetric for the command-triggered case. Unfortunately, the timing and the time stamp models proposed in Proposition \ref{prop:final_model} cannot easily be used together with a plant model to estimate jointly the plant state and the timing. However, the problems are not with the timing and the time stamp models as such but rather of more fundamental nature:
First of all, consider a time delay $\delta$ in transfer function form
\begin{equation*}
e^{\delta s}=1+\delta s+\frac{(\delta s)^2}{2\!}+\frac{(\delta s)^3}{3\!}+\dots.
\end{equation*}
The exponential is nonlinear but what is even worse is that the coupling between the plant and the time uncertainty cannot be modeled exactly with a finite number of states showing that the optimal filter is not realizable and some approximations have to be employed.
Secondly, the timing instants are not dynamic states in the usual sense. The state transition from one to the next timing instants are dependent on the difference between the corresponding timing instants. Also the timing determines the order of the data. In other words, the topology of the estimation is dependent on previous timing instants, which means that they should continuously be added as states. Based on similar arguments the optimal filter in the linear system case is deemed nonrealizable in \citet{Thomopoulos1994a}. 
Thirdly, the effect of a timing error of a control command is dependent on a continuum of the control input which might not be easily accessible and probably will have to be approximated.





Fortunately, though the optimal filter is not realizable, it has not stopped the development of approximative methods as reviewed in Section~\ref{subsec:previous_work}. In~\citep{Philipp2010}, the presented estimation methods incorporate a structure similar to \eqref{eq:1}-\eqref{eq:4}. 
Also, the methods presented in~\citep{Julier2005,Nilsson2010a,Thomopoulos1994a,Shin1995} and some of the methods found in~\citep{Ferreira1997,Bjoerklund2003} could be used together with \eqref{eq:3}-\eqref{eq:2}. However, how to combine them with \eqref{eq:3}-\eqref{eq:2} is not obvious. These methods assume that the timing error is white, constant, or possesses limited structure, e.g., constant plus white noise. However, they have a common important characteristic in that they \emph{give the state and the timing posterior estimates given a measurement and a timing prior}. In a sense, going from the timing prior to the posterior estimate can be seen as estimating the timing error. This timing error in turn are dependent on estimation errors of the remaining states in \eqref{eq:3}-\eqref{eq:2}. Therefore, together with complementary filtering, these methods could provide the link between state estimation and timing estimation. A timing prior and its coupling to the remainder of the timing states would be given by \eqref{eq:3}-\eqref{eq:2} for an incoming measurement or control command and subsequently all states of \eqref{eq:3}-\eqref{eq:2} would be updated with incoming measurements by these methods. For event-triggered sampling, the additional information given by the state constraint of the plant must also be handled in the update phase. The timing order problem would probably most naturally be solved as suggested in~\citep{Thomopoulos1994a}, by fixing the timing estimate to the posterior estimate. For control purposes this method would combine well with methods presented in \citet{Nilsson1998} because the methods assume \emph{a posteriori} perfectly known time instants. Even though this is not the case, fixing the time instant estimates to the \emph{a posteriori} estimates will mimic this.



\section{Jitter reduction}\label{sec:redu_jitter}
Even though the clock pace model \eqref{eq:2} and the communication delay model \eqref{eq:3} do not need to be very complicated, \eqref{eq:3}-\eqref{eq:2} still add complexity. Further, the integration with plant state estimation and control algorithms as outlined in Section~\ref{sec:heuristics} is not necessarily trivial. Hence the timing estimation needs to be justified with increased performance.

For scenario 1 any time variations in the communication delay (apart from trends possibly tracked in \eqref{eq:3}) will penetrate through as a perceived jitter in the timing prior.
In general the reduction in jitter in scenario 2 in comparison with scenario 1 is difficult to assess.
However, for an illustrative simple case the gain can be derived in close-form solution and a qualitative description of the jitter reduction can be achieved through continuity arguments:
\begin{prop}\label{thm:redu}
If $\alpha^i_{k|r}$ and $\delta^i_k$ are constants plus white Gaussian noise with standard deviation (std) $\sigma_\alpha$ and $\sigma_\delta$ respectively, $q^i_k$ is constant, and reports arrive with unit intervals $d^{i}_{k}=1$, the steady-state minimum-mean-square error (mmse) timing prior error std for scenario 1 is $\sigma_\delta$ and for scenario 2
\begin{equation}\label{eq:a_posteriori_var}
\sigma_\delta-\frac{\sigma_\delta^2}{\sigma_\delta+\frac{1}{2}\sigma_\alpha+\frac{1}{2}\left(\sigma_\alpha^2+4\sigma_\delta \sigma_\alpha\right)^{\nicefrac{1}{2}}}.
\end{equation}
\end{prop}

At steady-state the estimation of the constant components of $\alpha_{k,r|c}$ and $\delta^i_k$ must have converged. For scenario 1, from Proposition \ref{prop:final_model} it is easily seen that the timing prior standard deviation is $\sigma_\delta$. For scenario 2, Proposition \ref{prop:final_model} gives a linear system with state, output, and process and measurement noise covariance matrices
\begin{equation*}
\begin{bmatrix}
1 & 1 \\
0 & 1
\end{bmatrix},
\quad
\begin{bmatrix}
1\\
0
\end{bmatrix},
\quad
\begin{bmatrix}
\sigma_\alpha^2 & 0\\
0 & 0
\end{bmatrix},
\quad\text{and}\quad
\begin{bmatrix}
\sigma_\delta^2
\end{bmatrix}.
\end{equation*}
Solving the steady-state Riccati equation gives the mmse \emph{a posteriori} error standard deviation \eqref{eq:a_posteriori_var} which is the timing prior standard deviation.

The point is that obviously
\begin{equation}\label{eq:gain}
\sigma_\delta-\frac{\sigma_\delta^2}{\sigma_\delta+\frac{1}{2}\sigma_\alpha+\frac{1}{2}\left(\sigma_\alpha^2+4\sigma_\delta \sigma_\alpha\right)^{\nicefrac{1}{2}}}\leq \sigma_\delta.
\end{equation}
The effect is especially prominent when $\sigma_\alpha\ll \sigma_\delta$ which implies the same relation for \eqref{eq:gain}.
Due to the continuity of the Riccati equations one would expect similar behavior even if $\alpha^{-1}_{k,i|c}$ is not a constant but $|\alpha^{-1}_{k+1,i|c}-\alpha^{-1}_{k,i|c}|\ll \sigma_\delta$, $|\alpha^{-1}_{k+1,i|c}-\alpha^{-1}_{k,i|c}|\ll d^i_k$, and if there is sufficient plant dynamics. Both $|\alpha^{-1}_{k+1,i|c}-\alpha^{-1}_{k,i|c}|\ll \sigma_\delta$ and $|\alpha^{-1}_{k+1,i|c}-\alpha^{-1}_{k,i|c}|\ll d^i_k$ can be expected for most clocks and communication links. Consequently using peripheral unit time stamps in addition to those from the central unit can greatly reduce jitter and therefore improve performance of sensor and control systems. By using peripheral unit time stamps the jitter is essentially determined by the peripheral unit clock stability. Consequently even rather crude communication delay models can be expected to work.

Note that the models assumptions of Proposition \ref{thm:redu} are similar to those used in most clock synchronization schemas and are therefore reasonable for many systems.

\begin{figure}[t!]
\centering
\includegraphics[width=\linewidth]{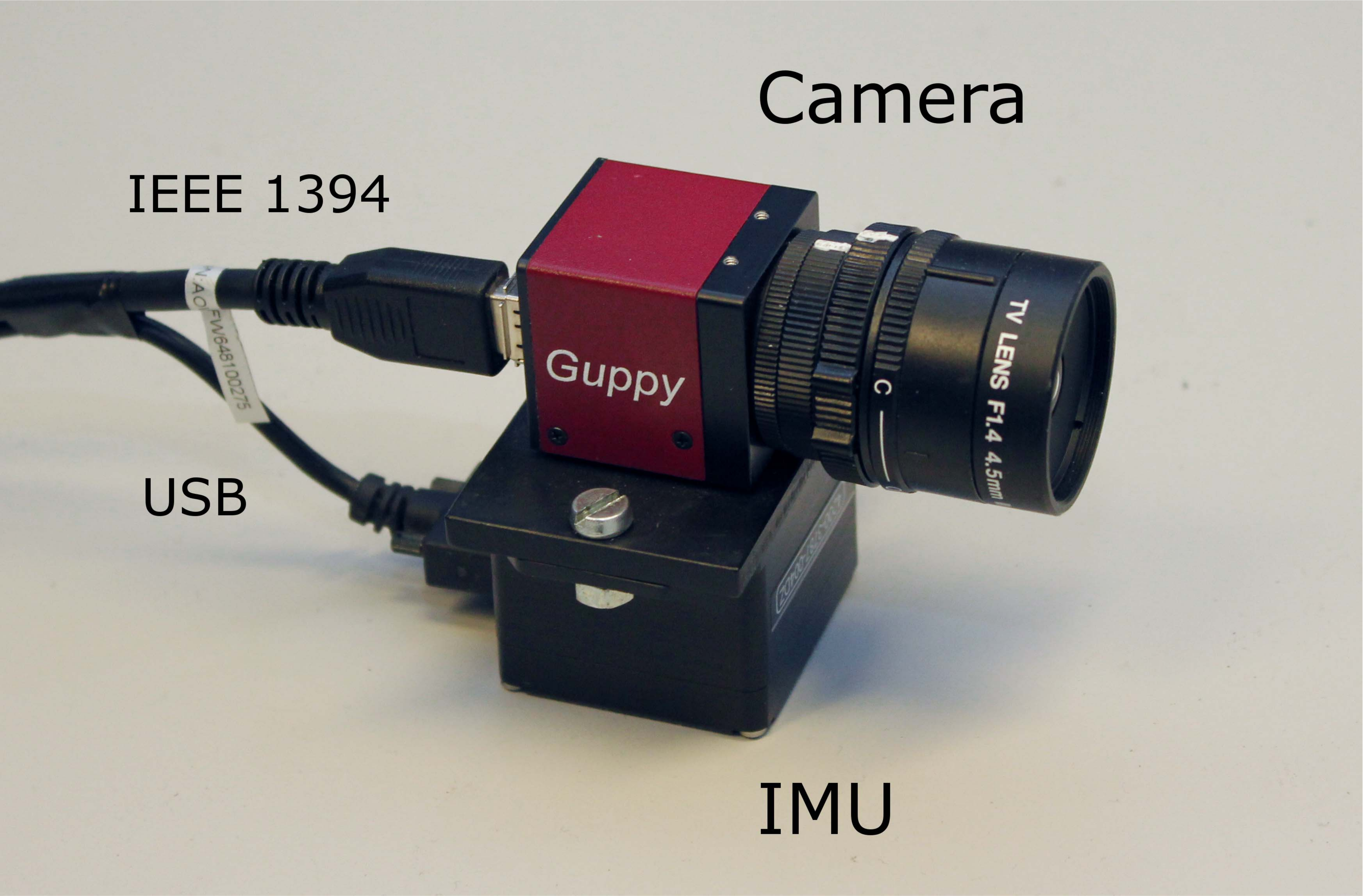}
\caption{Prototyping setup for camera aided inertial navigation. A camera is mounted together with an IMU and each sensor unit is connected to a processing device with asynchronous data busses, i.e., IEEE-1394 and USB.}\label{fig:cam_setup}
\end{figure}

\subsection{Numerical example}\label{subsec:example}
A camera and an inertial measurement unit (IMU) is mounted together. In a prototyping setup, the combination is used for camera-aided inertial navigation (see for example \citet{Zachariah2011}). The camera and the IMU are connected with asynchronous data busses to a computer used for processing the sensor data. An illustration of the setup is found in Figure~\ref{fig:cam_setup}. Different threads are used to receive the data and the data are time stamped in application code as soon as a thread is scheduled to take care of the incoming data.  The image processing is computationally heavy and therefore the CPU load is expected to be high. For simplicity, interoperability with other applications, and low overhead no real-time functionality is used. Both the camera and the gyroscope are time-triggered. The data busses are only used for the navigation sensors and therefore the load is constant and models assumption as of Proposition~\ref{thm:redu} are made with $\sigma_\delta=5\cdot 10^{-3}[s]$ (half of typical time slice for unprioritized scheduling) and $\sigma_\alpha=10^{-6}[s]$ (period of peripheral oscillator). This gives for scenario 1 a jitter std of $5\cdot 10^{-3}[s]$ and for scenario 2 a jitter std of $3\cdot 10^{-5}[s]$. For a panning motion of $90[{}^\circ/s]$ and a camera horizontal angular resolution of $10[pixel/{}^\circ]$ (resolution $640\times460$ and $46^\circ$ horizontal field of view) this jitter would induce a discrepancy of the predicted view rotation to the perceived view rotation with std of $4.5[pixel]$ for scenario 1 and $0.03[pixel]$ for scenario 2. With the discrepancy of scenario 1 the navigation performance would probably deteriorate significantly. This example shows that by including peripheral unit timing information one can do without either off-line preprocessing, driver programming, real-time functionality, or thread priority tweaking, which will simplify the prototyping significantly.

\section{Summary and conclusions}\label{sec:conclusions}
This article analyzes the timing of a distributed sensor or control system with central processing. Measurement models for the time stamps have been derived from underlying clock and communication delay definitions and a heuristically defined system time. Communication delay and clock skew models have been argued to be necessary to model the timing and based on such models and time stamps measurement models a complete model of the timing, and its relation to the time stamps, has been proposed. It has been sketched how the coupling with the plant model can be handled. Finally, based on a closed-form solution for simple underlying models and continuity argument, the conclusion is that peripheral unit time stamps and explicit treatment of the timing estimation can give significant jitter reduction in comparison with that of a naive approach using only central unit time stamps.

Jitter in general lead to reduced performance and possibly even instability for control systems. Therefore explicit treatment of the timing estimation with models of Proposition~\ref{prop:final_model} is beneficial for any distributed sensor or control system, and possibly even necessary, for control systems with high plant dynamics or large delay variations, if clock synchronization cannot be implemented.


\bibliographystyle{plainnat}
{\scriptsize
\bibliography{timing_est}
}
\end{document}